\documentclass[prl, showpacs,preprintnumbers,amsmath,amssymb]{revtex4}

\usepackage{graphicx}
\usepackage{dcolumn}
\usepackage{bm}

\begin{document}

\title{Trace Detection of Metastable Helium Molecules \\ in Superfluid 
Helium by Laser-Induced Fluorescence}

\author{D.\,N.~McKinsey}
\email{daniel.mckinsey@yale.edu}
\author{W.\,H.~Lippincott}
\author{J.\,A.~Nikkel}
\author{W.\,G.~Rellergert}
\affiliation{Department of Physics, Yale University, New Haven, CT 06511}

\date{\today}

\begin{abstract}

    We describe an approach to detecting ionizing radiation that
    combines the special properties of superfluid helium with the
    sensitivity of quantum optics techniques.  Ionization in liquid
    helium results in the copious production of metastable $\rm
    He_{2}$ molecules, which can be detected by laser-induced
    fluorescence.  Each molecule can be probed many times using a
    cycling transition, resulting in the detection of individual
    molecules with high signal to noise.  This technique could be used
    to detect neutrinos, weakly interacting massive particles, and
    ultracold neutrons, and to image superfluid flow in liquid
    $^{4}$He.
 
\end{abstract}


\maketitle

\twocolumngrid

Renewed interest in superfluid helium as a particle detection medium
has been spurred by proposals in low-background particle astrophysics
(where superfluid helium has the advantage of extremely high
radiopurity\cite{Lan87,McK00}) and tests of the Standard Model using
ultracold neutrons (where superfluid helium is used as an ultracold
neutron production and storage medium\cite{Gol94,Doy94}).  This paper
introduces an evolution of these experiments by adding a component of
direct triplet He$_2$ molecule detection through laser-induced
fluorescence.

Experiments in the 1950s and 1960s showed that superfluid helium
scintillates brightly in the extreme ultraviolet when exposed to
ionizing radiation\cite{Fle59}, and that large numbers of
long-lived excitations are also created\cite{Sur68}.  Detailed
spectroscopy of electron-excited superfluid helium\cite{Den69} later
showed that both of these effects are due to the efficient production
of He$_{2}$ excimer molecules, particularly in the $\rm
He_{2}(\mathit{A}^{1} \Sigma^{+}_{\mathit{u}})$ and $\rm
He_{2}(\mathit{a}^{3}\Sigma^{+}_{\mathit{u}})$ states which produce
the scintillation and long-lived excitations respectively.  It has
been determined that more than 50\% of the energy of an energetic
electron in liquid helium is converted into chemical energy in the
form of He$_{2}$ molecules, creating in total about 13000 $\rm
He_{2}(\mathit{a}^{3}\Sigma^{+}_{\mathit{u}})$ and 19000 $\rm
He_{2}(\mathit{A}^{1} \Sigma^{+}_{\mathit{u}})$ states per
MeV\cite{Ada01}.  ÊElectronically excited helium atoms are unstable in
liquid helium, rapidly reacting with ground-state helium atoms.  ÊFor
example, atoms in the $2^3$S state, which have a radiative lifetime of
8000 seconds in vacuum\cite{Woo75}, only survive 15 $\mu$s in the
liquid due to non-radiative bonding with ground state atoms to form
$\rm He_{2}(\mathit{a}^{3}\Sigma^{+}_{\mathit{u}})$
molecules\cite{Ket74a}.

While the lowest-energy singlet and triplet molecules both emit 80 nm
photons when they radiatively decay\cite{Sur70}, their
radiative lifetimes are markedly different.  ÊThe $\rm
He_{2}(\mathit{A}^{1} \Sigma^{+}_{\mathit{u}})$ has a lifetime of
about 1 ns \cite{Hil89}, while the $\rm
He_{2}(\mathit{a}^{3}\Sigma^{+}_{\mathit{u}})$ has a lifetime of
$13\pm 2$ s \cite{McK99} in superfluid helium and forms a bubble of
radius 5.3 $\rm \AA$ \cite{Hic71}.ÊThe photon-emitting transition from
$\rm He_{2}(\mathit{a}^{3}\Sigma^{+}_{\mathit{u}})$ to the
dissociative ground state, $\rm
He_{2}(\mathit{X}^{1}\Sigma^{+}_{\mathit{g}})$, is forbidden because
it requires a spin flip.  ÊA theoretical value of $18$ s for the
lifetime of $\rm He_{2}(\mathit{a}^{3}\Sigma^{+}_{\mathit{u}})$ in the
lowest vibrational state has been obtained by assuming the spin flip
is driven by spin-orbit (SO) coupling and treating the Breit-Pauli
spin-orbit Hamiltonian using first order perturbation theory
\cite{Cha89}.  The lifetimes of the lowest-energy triplet molecular
states in neon, argon, krypton, and xenon are much shorter, measured to be 6.6
$\mu$s, 3.2 $\mu$s, 350 ns, and 50 ns respectively\cite{Oka74}.  The
large difference between the lifetimes of helium and the heavier noble
gases is qualitatively due to the respective strengths of the SO
coupling, which scales roughly as Z$^4$.  In addition, the lowest
energy excited helium atom is in an $S$ state, while the excited neon,
argon, krypton, and xenon atoms are in $P$ states.  Therefore, the heavier
noble gas molecules contain an intrinsic orbital angular momentum to
contribute to the SO coupling that is lacking in helium molecules.

\begin{figure}[!hbt] 
\includegraphics[width=3in]{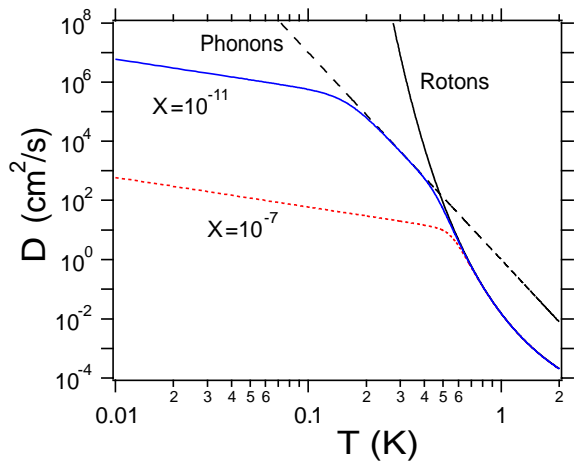} 
\caption{Diffusion constant $D$ versus temperature $T$ for He$_{2}$
molecules in superfluid helium, assuming He$^3$ concentrations found
in commercially available liquid helium ($X=10^{-7}$) and purified
liquid helium ($X=10^{-11}$).  Also shown are the individual roton and
phonon contributions.}
\label{fig:Diff} 
\end{figure}

 The metastable $\rm He_{2}(\mathit{a}^{3}\Sigma^{+}_{\mathit{u}})$
 molecule can also decay through reactions with other helium
 molecules\cite{Ket74a} and collisions with container walls.  ÊThe
 movement of the molecule is limited by diffusive scattering with
 rotons, phonons and $^3$He impurities (see Figure~\ref{fig:Diff}). 
 ÊWe model the diffusion constant by the following equation:
\begin{equation} 
\frac{1}{D}=\frac{e^{-\Delta/T}}{\Gamma_r}+\frac{T^7}{\Gamma_p}+X\frac{T}{\Gamma_3} 
\end{equation} 
where $X$ is the relative concentration of $^3$He in $^4$He, and
$\Delta$ is the roton energy gap (8.6 K).  $\Gamma_r$, $\Gamma_p$, and
$\Gamma_3$ represent measures of the scattering of the $\rm
He_{2}(\mathit{a}^{3}\Sigma^{+}_{\mathit{u}})$ bubble with rotons,
phonons, and $^3$He.  $\Gamma_r$ and $\Gamma_3$ have been determined
experimentally \cite{Rob72}, while $\Gamma_p$ has been scaled
from an experimental value for phonon-$^3$He scattering by taking into
account the greater effective mass of the He$_2$
bubble\cite{Hic71,Lam01}.  By adjusting the temperature and
$^3$He concentration, the displacement of $\rm
He_{2}(\mathit{a}^{3}\Sigma^{+}_{\mathit{u}})$ over its radiative
lifetime can be controlled, ranging from $0.7$ cm to $40$ m between 1
K and 100 mK.

\begin{figure}[!hbt] 
\includegraphics[width=3in]{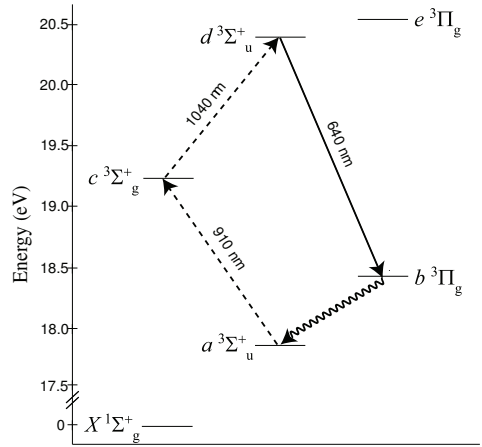} 
\caption{Energy levels of the triplet $\rm He_{2}$ molecule below 21 eV. Also 
shown is the cycling transition described in the text.} 
\label{fig:energylevels} 
\end{figure}

Many detailed spectroscopic studies of triplet $\rm He_{2}$ molecules
in superfluid helium have been performed, and the energy levels and
optical absorption frequencies are well
established\cite{Den69,Ket74a,Elt95}.  Figure
\ref{fig:energylevels} shows an energy level diagram of the lowest
lying triplet states of the $\rm He_{2}$ molecule.  ÊThe arrows on the
diagram show a cycling transition that can be used to detect $\rm
He_{2}$ molecules through laser-induced fluorescence.  The scheme
involves two pulsed infrared lasers.  ÊThe first laser excites the
molecule from $\rm \mathit{a}^{3}\Sigma^{+}_{\mathit{u}} \rightarrow
\rm \mathit{c}^{3}\Sigma^{+}_{\mathit{g}}$ with a pulse of 910 nm
light.  Immediately following is a second pulse to excite the
molecules from $\rm \mathit{c}^{3}\Sigma^{+}_{\mathit{g}} \rightarrow
\rm \mathit{d}^{3}\Sigma^{+}_{\mathit{u}}$.  Measured superfluid
helium emission spectra indicate that this transition is at 1040
nm\cite{Ket74a}, though the absorption wavelength is expected to be
blue-shifted slightly from this value\cite{Den69}.  Molecules in the
$\rm \mathit{d}^{3}\Sigma^{+}_{\mathit{u}}$ then decay to the $\rm
\mathit{b}^{3}\Pi_{\mathit{g}}$ with a 90\% branching ratio, emitting
an easily detected photon at 640 nm.  ÊThe lifetime of the $\rm
\mathit{d}^{3}\Sigma^{+}_{\mathit{u}} \rightarrow \rm
\mathit{b}^{3}\Pi_{\mathit{g}}$ transition has been measured to be 25
ns in both liquid\cite{Ben99} and gaseous\cite{Nee94} helium,
demonstrating that there is little non-radiative quenching of the $\rm
\mathit{d}^{3}\Sigma^{+}_{\mathit{u}}$ state.  Finally, the $\rm
\mathit{b}^{3}\Pi_{\mathit{g}}$ state is non-radiatively quenched
before it can decay\cite{Den69}, returning to the $\rm
\mathit{a}^{3}\Sigma^{+}_{\mathit{u}}$ state where it is again
sensitive to the probe laser.  Thus a single triplet $\rm He_{2}$
molecule is capable of emitting as many as $4\times 10^{7}$ photons/s. 
ÊGiven that the cross section of the $\rm
\mathit{a}^{3}\Sigma^{+}_{\mathit{u}} \rightarrow \rm
\mathit{c}^{3}\Sigma^{+}_{\mathit{g}}$ transition in superfluid helium
is $\rm 2.4\times 10^{-15}\, cm^{2}$\cite{Elt95} and assuming a
comparable cross section for the $\rm
\mathit{c}^{3}\Sigma^{+}_{\mathit{g}} \rightarrow \rm
\mathit{d}^{3}\Sigma^{+}_{\mathit{u}}$ transition, lasers emitting 250
$\rm \mu J/cm^{2}$ pulses at each wavelength would drive the $\rm
\mathit{a}^{3}\Sigma^{+}_{\mathit{u}} \rightarrow \rm
\mathit{d}^{3}\Sigma^{+}_{\mathit{u}}$ transition with roughly 90\%
efficiency per molecule. The linewidth requirement for the pump 
lasers is not stringent because the transitions are broadened by about 
120 cm$^{-1}$ by the liquid helium\cite{Elt95}. 

Benderskii \textit{et al.} have demonstrated that optical excitation
at 462 nm can drive the $\rm \mathit{a}^{3}\Sigma^{+}_{\mathit{u}}
\rightarrow \rm \mathit{e}^{3}\Pi_{\mathit{g}}$ transition in
superfluid helium.  ÊThe $\rm \mathit{e}^{3}\Pi_{\mathit{g}}$ state
then has a high probability of non-radiatively decaying to the $\rm
\mathit{d}^{3}\Sigma^{+}_{\mathit{u}}$ state\cite{Ben99}.  ÊThis gives
another scheme for driving the molecules to the $\rm
\mathit{d}^{3}\Sigma^{+}_{\mathit{u}}$ state, also resulting in the
emission of 640 nm fluorescence.  An advantage of this scheme is that
only one laser is used.  However, the blue excitation light is more
likely than infrared light to induce spurious fluorescence in any
associated optical elements, which could interfere with the detection
of 640 nm $\rm \mathit{d}^{3}\Sigma^{+}_{\mathit{u}} \rightarrow \rm
\mathit{b}^{3}\Pi_{\mathit{g}}$ fluorescence.

When driving a cyclic transition in molecules, complications can arise
from vibrational and rotational structure; if a large signal per
molecule is needed, it would be undesirable to have molecules falling
to a state in which they are not sensitive to the optical excitation
frequency.  Eltsov \textit{et al.} have measured a vibrational
relaxation time of 140 $\pm$ 40 ms for $\rm
\mathit{a}^{3}\Sigma^{+}_{\mathit{u}}$ molecules in superfluid helium,
and they demonstrated that the rotational relaxation is much faster
\cite{Elt95}.  From calculation of Franck-Condon factors for the
cycling transition shown in Figure~\ref{fig:energylevels}, we have
determined that only 1.2\% of the molecules per cycle will fall
radiatively to the first vibrational state rather than the zeroth
vibrational state.  If an experiment using this technique required the
molecules to be cycled faster than the maximum effective vibrational
relaxation rate of 600 $\rm s^{-1}$, they can be repumped with a third
laser operating at 1070 nm that would drive the molecules from $a$(1)
to $c$(0) where they will decay to the $a$(0) approximately 95\% of
the time.  This repumping of the molecules into the ground vibrational
state can be used to ensure that the molecules are sensitive to the
optical frequency chosen for the cycling transition.  In other
experiments, it may be desirable to cause molecules created by one
ionizing radiation event to be undetectable during a following event. 
In this case, the molecules can be driven into a different vibrational
state where they will be blind to the pump lasers.

A detector using individual helium molecule fluorescence in a cycling
transition would be far more sensitive to small energy depositions
than a comparable scintillation detector.  ÊIt should be possible to
obtain usable signal from every molecule formed in the detection
region, whereas approaches using only prompt scintillation detection
have to contend with incomplete detector coverage, dark counts, and
the inherently low quantum efficiency of photomultipliers (typically
no better than 30\%).  Detection of individual molecules should allow
an energy threshold on the order of 100 eV for electron-like events,
assuming that molecule production is linear with energy deposition. 
Ionization in noble gases has been shown to be roughly linear with
energy deposition down to the 100 eV scale in neon and xenon
gas\cite{San91}, and we expect this also to be the case in helium. 
Use of a cycling transition allows a significant signal amplification;
the total photon production is roughly 13000 per MeV per cycle.  In
comparison, organic scintillator typically produces 10000 photons per
MeV.

In addition to having a very low energy threshold, a detector
incorporating laser-induced fluorescence for the detection of He$_{2}$
molecules can also have very good position resolution.  ÊBecause the
scheme shown in Figure~\ref{fig:energylevels} can only excite molecules
where the beams overlap, one can orient the two beams at 90 degrees to
one another and raster them throughout the volume of interest.  ÊThis
can lead to position resolution equal to or less than the radius of
the laser beam profile.  ÊAlternately, it should be possible to image
tracks in the superfluid helium using an image intensifying unit or
CCD.

With the properties mentioned above, a detector using superfluid
helium could be used in several different configurations.  One mode of
operation is a simple continuous scan of the detector volume to detect
energy deposition events.  ÊA second is to use a layer of wavelength
shifting fluor\cite{McK97} to convert the singlet $\rm He_{2}$
scintillation light to the visible.  The visible light would be
detected by photomultipliers, triggering the laser
scanning system.  The lasers need only be activated following event
detection, and some particle identification should be possible by
measuring the ratio of prompt scintillation light to laser-induced
fluorescence.  In other liquified noble gases, this ratio has been
found to vary with excitation type\cite{Aki02,Hit83}, which is likely
the case in liquid helium as well\cite{Ada01}.  In addition, requiring
the detection of both prompt scintillation light and laser-induced
fluorescence would allow the rejection of backgrounds due to any
interactions that do not take place within the liquid helium.  The
disadvantage of this second mode is that the energy threshold will be
set by the efficiency of detecting the prompt scintillation light,
while in the first mode the energy threshold is set by the amplified
triplet molecule signal.

 The projected low energy threshold provides a unique opportunity to
 search for the neutrino magnetic moment.  The differential
 cross-section for magnetic neutrino-electron elastic scattering is
 given by
 $$(\frac{d\sigma}{dT})_{\mathrm{MS}}=\frac{\pi\alpha^{2}_{\mathrm{em}}\mu_{\ell}^{2}}{m_{e}^{2}}[\frac{1-T/E_{\nu}}{T}]$$
 where $\alpha_{\mathrm{em}}$ is the fine structure constant,
 $m_{e}$ is the electron mass, $T$ is the kinetic energy of the
 recoil electron and $E_{\nu}$ is the neutrino energy\cite{Vog89}. 
 Because this scattering process is enhanced for low $T$, tighter
 limits on the neutrino magnetic moment $\mu_{\ell}$ can be set with a
 low-radioactivity detector that also has a low energy threshold.  A
 radiopure sample of superfluid helium, scanned by infrared lasers and
 viewed by photodetectors, could have an energy threshold as low as
 100 eV. In addition, the very good position resolution available with
 this technique would assist in the characterization of systematic
 uncertainties.  Gamma ray backgrounds from Compton scattering can be
 partially rejected by looking for multiple scattering events in the
 detector volume, and x-ray backgrounds can be reduced by considering
 only events that occur within the central region of the detector,
 where x-rays are unlikely to penetrate.  To avoid significant
 He$_{2}$ diffusion, and to avoid scattering of laser light from
 bubbles in the liquid, this detector would be maintained at a
 temperature slightly below the lambda point, at roughly $\rm 2~K$. 
 The low energy threshold projected for this detector would also allow
 the detection of coherent neutrino-nucleus
 scattering\cite{Dru84}.

Laser-induced fluorescence in superfluid helium may also be useful in
the search for dark matter in the form of weakly interacting massive
particles (WIMPs).  ÊThe HERON group has demonstrated~\cite{Ban95}
that alpha particle excitations can result in a directional roton
signal in superfluid helium.  This occurs because the scattering rate
of rotons within the track is much higher than the scattering rate of
rotons once they leave, and the rotons are more likely to be emitted
perpendicular to an elongated track than parallel to it.  We speculate
that the trajectories of triplet molecules emerging from a
spin-independent WIMP-nucleus scattering event in superfluid $^{4}$He
would also show this directional effect, and thus provide information
about the direction of the nuclear recoil track.  Like the rotons, the
diffusion of molecules within the track would be dominated by roton
scattering, and the triplet molecules should be preferentially emitted
perpendicular to the track.  The molecule trajectories could then be
determined by a laser tracking system, and the pattern of molecule
trajectories could be used to determine the direction of the initial
nuclear recoil.  While DRIFT~\cite{Mor03} is promising for achieving
directional sensitivity for WIMP-nucleus scattering events using a
low-pressure time projection chamber, superfluid helium would allow a
$10^{5}$ times larger number density.  Meyer and Sloan estimate that a
nuclear recoil in $^3$He (from a WIMP, for example) will result in
50\% of the electronic excitation that would be expected from an
electron-like event for a similar energy\cite{Mey97}.  Assuming the
same figure for $^4$He, a nuclear recoil in superfluid helium will
form $6.5$ $\rm He_{2}(\mathit{a}^{3}\Sigma^{+}_{\mathit{u}})$
molecules per keV, allowing a very low energy threshold.  The same
detector, filled with $^{3}$He instead of $^{4}$He, could also be used
to place limits on spin-dependent WIMP scattering, though in this case
the much lower diffusion constant would preclude determination of the
nuclear recoil direction because of scattering of the molecule after 
it leaves the track. 

Laser-induced fluorescence may also be used to detect ultracold
neutrons in superfluid helium.  In a currently proposed search for the
neutron electric dipole moment (EDM)~\cite{Gol94}, He$_{2}$ singlet
light emission is to be used to detect neutron absorption events in
$^{3}$He-doped superfluid $^{4}$He.  These events in turn monitor the
precession of spin-polarized neutrons.  With the addition of
laser-induced fluorescence, event by event detection of both the
singlet and triplet molecules and measurement of the ratio of the
amplitudes of these two signals would allow the rejection of
backgrounds associated with gamma ray scattering in the liquid helium
and light collection optics.  Due to the low temperature of the
superfluid helium in the planned neutron EDM cells, the large He$_{2}$
diffusion constant will cause any triplet molecules to be quenched
by collisions with the wall within about 3 ms.  This
is adequate time to cycle the triplet molecules multiple times
with little effect from previous events, provided that the event rate
is less than $\sim$ 100 $\rm s^{-1}$ .  The same detection technique
could also be used in experiments to measure the neutron lifetime
using magnetically trapped neutrons\cite{Doy94}.

A fascinating non-particle physics application of laser-induced
fluorescence is the study of superfluid flow and turbulence.  ÊA small
radioactive source, focused laser beam, or electric discharge could be
used to create triplet He$_{2}$ molecules that would then be tracked
with intersecting lasers to image their path.  ÊOne approach to
imaging superfluid flow is neutron absorption
tomography~\cite{Hayden04}, which uses $^{3}$He as a neutral tracer
and requires a finely collimated neutron beam and the ability to
raster the neutron beam through the region of interest.  ÊBy instead
using laser induced fluorescence of $\rm
He_{2}(\mathit{a}^{3}\Sigma^{+}_{\mathit{u}})$ molecules, many of the
same superfluid helium properties can be measured, with the advantages
of better position resolution and imaging in three dimensions instead
of only two.  Another method of imaging superfluid flow is 
particle image velocimetry with small (1-10 $\rm \mu m$) neutrally
buoyant glass beads\cite{Don02}, but this approach has so far been
limited to temperatures above the lambda point.

In summary, we propose laser-induced fluorescence of metastable 
He$_{2}$ molecules as a technique for the detection of ionizing 
radiation with low energy threshold and good position resolution. This 
appears useful for applications in neutrino physics, the search for 
WIMPs, ultracold neutron research, and the imaging of superfluid 
turbulence.  

 We thank R.\,L.\,Brooks, D.\,Budker, D.\,DeMille, and M.\,E.\,Hayden for
 helpful discussions.  This work was supported by the David
 and Lucile Packard Foundation.

\end{document}